\documentclass[manuscript]{acmart} 

\usepackage[utf8]{inputenc}
\usepackage[T1]{fontenc}
\usepackage{graphicx}
\usepackage{enumitem}
\usepackage{amsmath}
\usepackage{bbm}
\usepackage{tabularx}
\DeclareMathOperator*{\argmax}{argmax}

\newcommand {\otoprule }{\midrule[\heavyrulewidth]}

\def\BibTeX{{\rm B\kern-.05em{\sc i\kern-.025em b}\kern-.08emT\kern-.1667em\lower.7ex\hbox{E}\kern-.125emX}}
    
\setcopyright{none}
\acmConference[ComplexRec 2020]{Workshop on Recommendation in Complex Scenarios at the ACM RecSys Conference on Recommender Systems (RecSys 2020)}{22 September 2020}{Rio de Janeiro, Brazil}
\acmDOI{}
\acmISBN{}
\copyrightyear{2020. Copyright for the individual papers remains with the authors. Copying permitted for private and academic purposes. This volume is published and copyrighted by its editors}
\acmPrice{}
\begin{document}

%
\title[CuratorNet: Visually-aware Recommendation of Art Images]{CuratorNet: Visually-aware Recommendation of Art Images}

%

\author{Pablo Messina}
\affiliation{%
  \institution{Pontificia Universidad Cat\'olica}
  \city{Santiago}
  \country{Chile}
}
\email{pamessina@uc.cl}
\additionalaffiliation{%
  \institution{Millennium Institute Foundational Research on Data, IMFD}
  \city{Santiago}
  \country{Chile}
}

\author{Manuel Cartagena}
\affiliation{%
  \institution{Pontificia Universidad Cat\'olica}
  \city{Santiago}
  \country{Chile}
}
\email{micartagena@uc.cl}

\author{Patricio Cerda}
\affiliation{%
  \institution{Pontificia Universidad Cat\'olica}
  \city{Santiago}
  \country{Chile}
}
\email{pcerdam@uc.cl}

\author{Felipe del Rio}
\affiliation{%
  \institution{Pontificia Universidad Cat\'olica}
  \city{Santiago}
  \country{Chile}
}
\email{fidelrio@uc.cl}
\additionalaffiliation{%
  \institution{Millennium Institute Foundational Research on Data, IMFD}
  \city{Santiago}
  \country{Chile}
}

\author{Denis Parra}
\orcid{0000-0001-9878-8761}
\affiliation{%
  \institution{Pontificia Universidad Cat\'olica}
  \city{Santiago}
  \country{Chile}
}
\email{dparra@ing.puc.cl}
\additionalaffiliation{%
  \institution{Millennium Institute Foundational Research on Data, IMFD}
  \city{Santiago}
  \country{Chile}
}

\renewcommand{\shortauthors}{Cartagena, et al.}

%
\begin{abstract}

Although there are several visually-aware recommendation models in domains like fashion or even movies, the art domain lacks the same level of research attention, despite the recent growth of the online artwork market. To reduce this gap, in this article we introduce CuratorNet, a neural network architecture for visually-aware recommendation of art images.  CuratorNet is designed at the core with the goal of maximizing generalization: the network has a fixed set of parameters that only need to be trained once, and thereafter the model is able to generalize to new users or items never seen before, without further training. This is achieved by leveraging visual content: items are mapped to item vectors through visual embeddings, and users are mapped to user vectors by aggregating the visual content of items they have consumed. 
Besides the model architecture, we also introduce novel triplet sampling strategies to build a training set for rank learning in the art domain, resulting in more effective learning than naive random sampling. With an evaluation over a real-world dataset of physical paintings, we show that CuratorNet achieves the best performance among several baselines, including the state-of-the-art model VBPR. CuratorNet is motivated and evaluated in the art domain, but its architecture and training scheme could be adapted to recommend images in other areas.


\end{abstract}

%
%


%

%

%
\maketitle

\section{Introduction}


The big revolution of deep convolutional neural networks (CNN) in the area of computer vision for tasks such as image classification \cite{krizhevsky2012imagenet,simonyan2014very,he2016deep}, object recognition \cite{akcai2016},  image segmentation \cite{badrinarayanan2017segnet} or scene identification \cite{sharif2014cnn} has  reached the area of image recommender systems in recent years \cite{mcauley2015image,he2016vbpr,he2016vista,lei2016comparative,kang2017dvbpr,messina2018content}. These works use neural visual embeddings to improve the recommendation performance compared to previous approaches for image recommendation based on ratings and text \cite{aroyo2007personalized}, social tags \cite{semeraro2012folksonomy}, context \cite{benouaret2015personalizing} and manually crafted visual features \cite{van2006multimedia}. Regarding application domains of recent image recommendation methods using neural visual embeddings, to the best of our knowledge most of them focus on fashion recommendation \cite{mcauley2015image,he2016vbpr,kang2017dvbpr}, a few on art recommendation \cite{he2016vista,messina2018content} and photo recommendation \cite{lei2016comparative}. He et al. \cite{he2016vista} proposed Vista, a model combining neural visual embeddings, collaborative filtering as well as temporal and social signals for digital art recommendation.


However, digital art projects can differ significantly from physical art (paintings and photographs). Messina et al. \cite{messina2018content} study recommendation of paintings in an online art store using a simple k-NN model based on neural visual features and metadata. Although memory-based models perform fairly well, model-based methods using neural visual features report  better performance \cite{he2016vbpr,he2016vista} in the fashion domain, indicating room for improvement in this area, considering the growing sales in the global online artwork market\footnote{\url{https://www.artsy.net/article/artsy-editorial-global-art-market-reached-674-billion-2018-6}}.


The most popular model-based method for image recommendation using neural visual embeddings is VBPR \cite{he2016vbpr}, a state-of-the-art model that integrates implicit feedback collaborative filtering with neural visual embeddings into a Bayesian Personalized Ranking (BPR) learning framework \cite{rendle2009bpr}. VBPR performs well, but it has some drawbacks. VBPR learns a latent embedding for each user and for each item, so new users cannot receive suggestions and new items cannot be recommended until re-training is carried out. 
An alternative is training a model such as Youtube's Deep Neural Recommender \cite{covington2016deep} which allows to recommend to new users with little preference feedback and without additional model training. However, Youtube's model was trained on millions of user transactions and with large amounts of profile and contextual data, so it does not easily fit  to datasets that are small, with little user feedback or with little contextual and profile data. 

In this work, we introduce a neural network for visually-aware recommendation of images focused on visual art named \emph{CuratorNet}, whose general structure can be seen in Figure \ref{fig:acnet-train}. CuratorNet leverages neural image embeddings as those obtained from CNNs  \cite{krizhevsky2012imagenet,simonyan2014very,he2016deep} pre-trained on the Imagenet dataset (ILSVRC \cite{ILSVRC15}). 
We train CuratorNet for ranking with triplets ($P_u$, $i_+$, $j_-$), where $P_u$ is the history of image preferences of a user $u$, whereas $i_+$ and $j_-$ are a pair of images with higher and lower preference respectively.
CuratorNet draws inspiration from VBPR \cite{he2016vbpr} and Youtube's Recommender System \cite{covington2016deep}. VBPR \cite{he2016vbpr} inspired us to leverage pre-trained image embeddings as well as optimizing the model for ranking as in BPR \cite{rendle2009bpr}. From the work of Convington et al. \cite{covington2016deep} we took the idea of designing a deep neural network that can generalize to new users without introducing new parameters or further training (unlike VBPR which needs to learn a latent user vector for each new user). As a result, CuratorNet can recommend to new users with very little feedback, and without additional training 
\textit{CuratorNet}'s deep neural network is trained for personalized ranking using triplets and the architecture contains a set of layers with shared weights, inspired by models using triplet loss for non-personalized image ranking \cite{schroff2015facenet,wang2014learning}. In these works, a single image represents the input query, but in our case, the input query is a set images representing a user preference history, $P_u$.  In summary, compared to previous works, our main contributions are: 
\begin{itemize}

    \item a novel neural-based visually-aware architecture for image recommendation, 
    \item  a set of sampling guidelines for the creation of the training dataset (triplets), which improve the performance of \textit{CuratorNet} as well as \textit{VBPR} with respect to random negative sampling, and
    \item  presenting a thorough evaluation of the method against competitive state-of-the-art methods (VisRank \cite{kang2017dvbpr,messina2018content} and VBPR\cite{he2016vbpr}) on a dataset of purchases of physical art (paintings and photographs).
\end{itemize}

We also share the dataset\footnote{\url{https://drive.google.com/drive/folders/1Dk7_BRNtN_IL8r64xAo6GdOYEycivtLy}} of user transactions (with hashed user and item IDs due to privacy requirements) as well as visual embeddings of the paintings image files. One aspect to highlight about this research, is that although the triplets' sampling guidelines to  build the BPR training set apply specifically to visual art, the architecture of \textit{CuratorNet} can be used in other visual domains for image recommendation.

\section{Related Work}
\label{sec:related-work}
In this section we provide an overview of relevant related work, considering:  \textit{Artwork Recommender Systems} (\ref{sec:artrecsys}),  \textit{Visually-aware Recommender Systems} (\ref{sec:visawrecsys}), as well as highlights of what differentiates our work to the existing literature.

\subsection{Artwork Recommender Systems}
\label{sec:artrecsys}

With respect to artwork recommender systems, one of the first contributions was the CHIP Project \cite{aroyo2007personalized}. The aim of the project was to build a recommendation system for the Rijksmuseum. The project used traditional techniques such as content-based filtering based on metadata provided by experts, as well as collaborative filtering based on users' ratings. Another similar system but non-personalized was $m4art$ by Van den Broek et al. \cite{van2006multimedia}, who used color histograms to retrieve similar
art images given a painting as input query.

Another important contribution is the work by Semeraro et al. \cite{semeraro2012folksonomy}, who introduced an artwork recommender system called FIRSt (Folksonomy-based Item Recommender syStem) which utilizes social tags given by experts and non-experts of over 65 paintings of the Vatican picture gallery. They did not employ visual features among their methods. Benouaret et al. \cite{benouaret2015personalizing} improved the state-of-the-art in artwork recommender systems using context obtained through a mobile application, with the aim of making museum tour recommendations more useful. Their content-based approach used ratings given by the users during the tour and metadata from the artworks rated, e.g. title or artist names. 

Finally, the most recent works use neural image embeddings \cite{he2016vista,messina2018content}. He et al. \cite{he2016vista} propose the system Vista, which addresses digital artwork recommendations based on pre-trained deep neural visual features, as well as temporal and social data. On the other hand, Messina et al. \cite{messina2018content} address the recommendation of one-of-a-kind physical paintings, comparing the performance of metadata, manually-curated visual features, and neural visual embeddings. Messina et al. \cite{messina2018content} recommend to users by computing a simple K-NN based similarity score among users' purchased paintings and the paintings in the dataset, a method that Kang et al. \cite{kang2017dvbpr} call \textit{VisRank}.

\subsection{Visually-aware Image Recommender Systems}
\label{sec:visawrecsys}

In this section we survey works using visual features to recommend images. We also cite a few works using visual information to recommend non-image items, but these are not too relevant for the present research.

Manually-engineered visual features extracted from images (texture, sharpness, brightness, etc.) have been used in several tasks for information filtering, such as retrieval \cite{rui1998relevance,la1998combining,van2006multimedia} and ranking \cite{sanpedro2009}. More recently, interesting results have been shown for the use of low-level handcrafted stylistic visual features automatically extracted from video frames for content-based video recommendation \cite{deldjoo2016content}. Even better results are obtained when both stylistic visual features and annotated metadata are combined in a hybrid recommender, as shown in the work of Elahi et al. \cite{Elahi:2017:ESG:3109859.3109908}. In a visually-aware setting not related to recommending images, Elsweiller et al. \cite{elsweiler2017exploiting} used manually-crafted attractiveness visual features  \cite{sanpedro2009}, in order to recommend healthy food recipes to users.

Another branch of visually-aware image recommender systems focuses on using neural embeddings to represent images \cite{he2016vbpr,he2016vista,lei2016comparative,kang2017dvbpr,messina2018content}. The computer vision community has a large track of successful systems based on neural networks for several tasks \cite{krizhevsky2012imagenet,simonyan2014very,he2016deep,akcai2016,badrinarayanan2017segnet,sharif2014cnn}. This trend started from the outstanding performance of the AlexNet \cite{krizhevsky2012imagenet} in the Imagenet Large Scale Visual Recognition challenge (ILSVRC \cite{ILSVRC15}), but the most notable implication is that the neural image embeddings have shown impressive performance for transfer learning, i.e., for tasks different from the original one \cite{kornblith2018transfer,delrio2018transfer}. Usually these neural image embeddings are obtained from CNN models such as AlexNet \cite{krizhevsky2012imagenet}, VGG \cite{simonyan2014very} and ResNet \cite{he2016deep}, among others. Motivated by these results, McAuley et al. \cite{mcauley2015image} introduced an image-based recommendation system based on styles and substitutes for clothing using visual embeddings pre-trained on a large-scale dataset obtained from Amazon.com. Later, He et al. \cite{he2016vbpr} went further in this line of research and introduced a visually-aware matrix factorization approach that incorporates
visual signals (from a pre-trained CNN) into predictors of people's opinions, called VBPR. Their training model is based on Bayesian Personalized Ranking
(BPR), a model previously introduced by Rendle et al. \cite{rendle2009bpr}.

The next work by He et al. \cite{he2016vista} deals with visually-aware digital art recommendation, building a model called Vista which combines ratings, temporal and social signals and visual features.

Another relevant work was the research by Lei et al. \cite{lei2016comparative} who introduced comparative deep learning for hybrid image recommendation. In this work, they use a siamese neural network architecture for making recommendations of images using user information (such as demographics and social tags) as well as images in pairs (one liked, one disliked) in order to build a ranking model. The approach is interesting, but they work with Flickr photos, not artwork images, and use social tags, not present in our problem setting. The work by Kang et al. \cite{kang2017dvbpr} expands VBPR but they focus on generating images using Generative adversarial networks \cite{goodfellow2014generative} rather than recommending, with an application in the fashion domain. Finally, Messina et al. \cite{messina2018content} was already mentioned, but we can add that their neural image embeddings outperformed other visual (manually-extracted) and metadata features for ranking, with the exception of the metadata given by user's favorite artist, which predicted even better than neural embeddings for top@k recommendation.

\subsection{Differences to Previous Research}
\label{sec:diffprevres}

Almost all the surveyed articles on artwork recommendation have in common that they used standard techniques such as collaborative filtering and content-based filtering, as well as manually-curated visual image features, but only the most recent works have exploited visual features extracted from CNNs \cite{he2016vista,messina2018content}. In comparison to these works, we introduce a model-based approach (unlike the memory-based VisRank method by Messina et al. \cite{messina2018content}) and which recommends to cold-start items and users without additional model training (unlike \cite{he2016vista}).
With regards to more general work on visually-aware image recommender systems, almost all of the surveyed articles have focused on tasks different from art recommendation, such as fashion recommendation \cite{mcauley2015image,he2016vbpr,kang2017dvbpr}, photo \cite{lei2016comparative} and video recommendation \cite{Elahi:2017:ESG:3109859.3109908}. Only Vista, the work by He et al. \cite{he2016vista}, resembles ours in terms of the topic (art recommendation) and the use of visual features. Unlike them, we evaluate our proposed method, CuratorNet, in a dataset of physical paintings and photographs, not only digital art. Moreover, Vista uses social and temporal metadata which we do not have and many other datasets might not have either.
Compared to all these previous research, and to the best of our knowledge, CuratorNet is the first architecture for image recommendation that takes advantage of shared weights in a triplet loss setting, an idea inspired by the results of Wang et al. \cite{wang2014learning} and Schroff et al. \cite{schroff2015facenet}, but here adapted to the personalized image recommendation domain.


\begin{table}[t!pb]
\begin{center}
\caption{Notation for CuratorNet.}
\label{tab:notations-train-set}
 \tiny
 \begin{tabularx}{1.0\linewidth}{lX}
 \otoprule
 \small\strut \textbf{Symbol} &\small\strut \textbf{Description} \\
  \midrule
  
  \small\strut $U, I$ & \small\strut user set, item set \\
  
  \small\strut $u$ & \small\strut a specific user \\
  
  \small\strut $i,j$ & \small\strut a specific item (resp.) \\
  
  
  \small\strut $i_+,j_-$ & \small\strut a positive item and negative item (resp.) \\
  
  
  \small\strut $I_u^+$ or $P_u$& \small\strut set of all items which the user $u$ has expressed a positive preference (full history) \\
  
  
  \small\strut $I_{u,k}^+$ & \small\strut set of all items  which the user $u$ has expressed a positive preference up to his $k$-th purchase basket (inclusive) \\
  
  \small\strut $P_{u,k}$ & \small\strut set of all items  which the user $u$ has expressed a positive preference in his $k$-th purchase basket \\
  
  
  
  
  
  

  \bottomrule
 \end{tabularx}
\end{center}
\end{table}

\section{CuratorNet}
\label{sec:recommendation-methods}

\subsection{Problem Formulation}

We approach the problem of recommending art images from user positive-only feedback (e.g., purchase history, likes, etc.) upon visual items (paintings, photographs, etc.). Let  $U$ and $I$ be the set of users and items in a dataset, respectively. We assume only one image per each single item $i \in I$. Considering either user purchases or likes, the set of items for which a user $u$ has expressed positive preference is defined as $I_{u}^{+}$. 
In this work, we considered purchases to be positive feedback from the user.
Our goal is to generate for each user $ u \in U$ a personalized ranked list of the items for which the user still have not expressed preference, i.e., for $I  \setminus I_{u}^{+}$.

\subsection{Preference Predictor}

The preference predictor in CuratorNet is inspired by VBPR \cite{he2016vbpr}, a state-of-the-art visual recommender model. 
%

%
However, \textbf{CuratorNet has some important differences}. First, we do not use non-visual latent factors, so we remove the traditional user and item non-visual latent embeddings
. Second, we do not learn a specific embedding per user such as VBPR, but we learn a joint model that, given a user's purchase/like history, it outputs a single embedding which can be used to rank unobserved artworks in the dataset, similar to YouTube's Deep Learning network \cite{covington2016deep}. Another important difference of VBPR with CuratorNet is that the former has a single matrix $\mathbf{E}$ to project a visual item embedding $f_i$ into the user latent space. In CuratorNet, we rather learn a neural network  $\Phi(\cdot)$ to perform that projection, which receives as input either a single image embedding $\mathbf{f}_i$ or a set of image embeddings representing users' purchase/like history $P_u = \{\mathbf{f}_1,...,\mathbf{f}_N\}$ . Given all these aspects, the preference predictor of CuratorNet is given by:

\begin{figure*}[t!]
    \centering
    \includegraphics[scale=0.42]{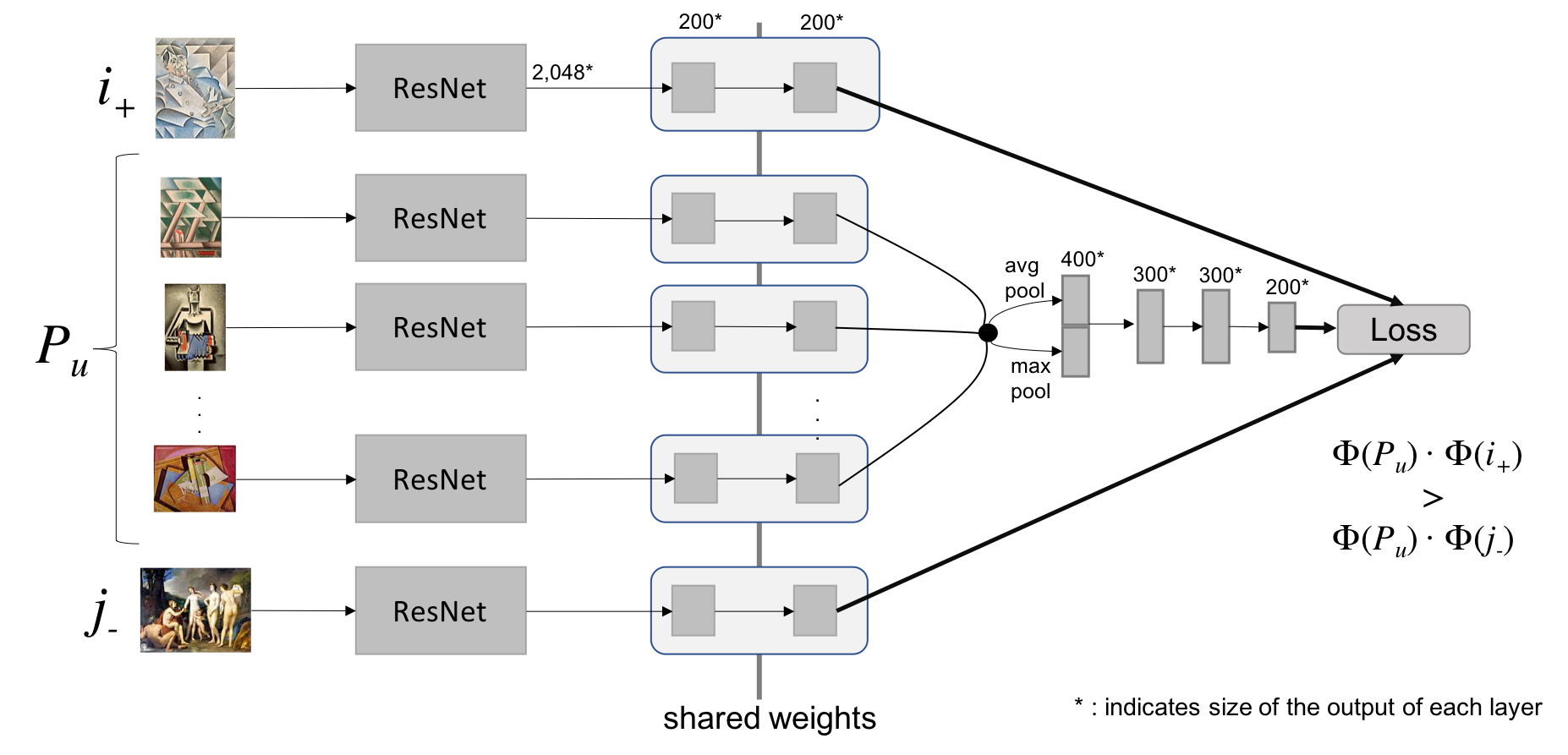}
    \caption{Architecture of CuratorNet showing in detail the layers with shared weights for training.}
    \vspace{-4mm}
    \label{fig:acnet-train}
\end{figure*}

\begin{equation}
    x_{u,i} = \alpha + \beta_u + \Phi(P_u)^T \Phi(\mathbf{f}_i) 
    \label{eq:CuratorNet-predictor}
\end{equation}

where $\alpha$ is an offset, $\beta_u$ represents a user bias, $\Phi(\cdot)$ represents CuratorNet neural network and $P_u$ represents the set of visual embeddings of the images in user $u$ history. After some experiments we found no differences between using or not a variable for item bias $\beta_i$ so we dropped it in order to decrease the number of parameters (Occam's razor). 

Finally, since we calculate the model parameters using BPR \cite{rendle2009bpr}, the parameters  $\alpha$, $\beta_u$ cancel out (details in the coming subsection) and our final preference predictor is simply

\begin{equation}
    x_{u,i} =  \Phi(P_u)^T \Phi(\mathbf{f}_i) 
    \label{eq:CuratorNet-predictor}
\end{equation}

\subsection{Model Learning via BPR}
\label{sec:model-learning}

We use the Bayesian Personalized Ranking (BPR) framework \cite{rendle2009bpr} to learn the model parameters. Our goal is to optimize ranking by training a model which orders triples of the form $(u, i, j) \in \mathcal{D_S}$, where $u$ denotes a user, $i$ an item with positive feedback from $u$, and $j$ an item with non-observed feedback from $u$. The training set of triples $\mathcal{D_S}$ is defined as: 

\begin{equation}
    \mathcal{D_S} = \{  (u,i,j)| u \in U \wedge i \in I_u^+ \wedge j  \in I \setminus I_u^+ \}
\end{equation}

Table \ref{tab:notations-train-set} shows that $I_+^u$ denotes the set of all items with positive feedback from $u$ while $I \setminus I_+^u$ shows those items without such positive feedback. Considering our previously defined preference predictor $x_{u,i}$, we would expect a larger preference score of $u$ over $i$ than over $j$, then BPR defines the difference between scores


\begin{equation}
    x_{u,i,j} =  x_{u,i} -  x_{u,j}
\end{equation}

an then BPR aims at finding the parameters $\Theta$ which optimize the objective function


\begin{equation}
    \label{eq:BPR}
    \argmax_\Theta \sum_{\mathcal{D_S}} \ln \sigma(x_{u,i,j}) - \lambda_{\Theta}||\Theta||^2
\end{equation}

where $\sigma(\cdot) $ is the sigmoid function, $ \Theta $ includes all model parameters, and $ \lambda_{\Theta} $ is a regularization hyperparameter.

In CuratorNet, unlike BPR-MF \cite{rendle2009bpr} and VBPR \cite{he2016vbpr}, we use a sigmoid cross entropy loss, considering that we can interpret the decision over triplets as a binary classification problem, where if $ x_{u,i,j} \textgreater 0 $ represents class $c=1$ (triple well ranked, since $ x_{u,i} \textgreater x_{u,j} $ ) and $ x_{u,i,j} \leq 0 $ signifies class $c=0$ (triplet wrongly ranked, since $ x_{u,i} \leq x_{u,j} $). Then, CuratorNet loss can be expressed as:


\begin{equation}
    \label{eq:cross-entropy}
    \mathcal{L} = -\sum_{ \mathcal{D_S}} c \ln( \sigma ( x_{u,i,j} ) ) + (1 - c) \ln( 1 - \sigma ( x_{u,i,j} ) ) + \lambda_{\Theta}||\Theta||^2
\end{equation}

where $c \in \{0,1\} $ is the class, $ \Theta $ includes all model parameters, $ \lambda_{\Theta} $ is a regularization hyperparameter, and  $\sigma( x_{u,i,j} ) $ is the probability that a user $u$ really prefers $i$ over $j$, $P(i >_u j | \Theta)$ \cite{rendle2009bpr},  calculated with the sigmoid function, i.e., 


\begin{equation}
    \label{eq:sigmoid}
    P(i >_u j | \Theta) = \sigma( x_{u,i,j} ) = \frac{1}{1 + e^{- (x_{u,i} -  x_{u,j}) }  }
\end{equation}

We perform the optimization to learn the parameters which reduce the loss function $\mathcal{L}$  by stochastic gradient descent with the Adam optimizer \cite{kingma2014adam}, using the implementation in Tensorflow\footnote{A reference CuratorNet implementation may be found at \url{https://github.com/ialab-puc/CuratorNet}.}. During each iteration of stochastic gradient descent, we sample a user $u$, a positive item $i \in I_+^u$ (i.e., removed from $P_u$), a negative item $j \in I \setminus I_+^u$, and user $u$  purchase/like history with item $i$ removed, i.e.,  $P_u \setminus i$.

\subsection{Model Architecture}
\label{sec:model-architecture}

The architecture of the CuratorNet neural network is summarized in Figure \ref{fig:acnet-train}, but is presented with more details in Figure \ref{fig:acnet-train}. For training, each imput instance is expected to be a triple ($P_u$,$i$,$j$), where $P_u$ is the set of images in user $u$ history (purchases, likes) with a single item $i$ removed from the set, $i$ is an item with positive preference, and  $j$ is an item with \textit{assumed} negative user preference. The negative user preference is assumed since the item $j$ is sampled from the list of images which $u$ has not interacted with yet. Each image ($i$, $j$ and all images $\in P_u$) goes through a ResNet \cite{he2016deep} (pre-trained with ImageNet data), which outputs a visual image embedding in $\mathbb{R}^{2,048}$. ResNet weights are fixed during CuratorNet's training. Then, the network has two layers with scale exponential linear units (hereinafter, SELU \cite{klambauer2017self}), with 200 neurons each, which reduce the dimensionality of each image. Notice that these two layers work similar to a siamese \cite{chopra2005learning} or triplet loss architecture \cite{wang2014learning,schroff2015facenet}, i.e., they have shared weights. Each image is represented at the output of this section of the network by a vector in $\mathbb{R}^{200}$. Then, for the case of the images in $P_u$, their embeddings are both averaged (average pooling \cite{boureau2010theoretical}) as well as max-pooled per dimension (max pooling \cite{boureau2010theoretical}) , and next concatenated to a resultant vector in $\mathbb{R}^{400}$. Finally, three SELU consecutive layers of 300, 200, and 200 neurons respectively end up with an output representation for $P_u$ in $\mathbb{R}^{200}$. The final part of the network is a ranking layer which evaluates a loss such that $ \Phi(P_u) \cdot \Phi(i) >  \Phi(P_u) \cdot \Phi(j) $, where replacing in Equation (\ref{eq:CuratorNet-predictor}), we have $ x_{u,i} >  x_{uj} $ . There are several options of loss functions, but due to good results of the cross-entropy loss in similar architectures with shared weights \cite{koch2015siamese} rather than, e.g. the hinge loss where we need to optimize an additional margin parameter $m$, we chose the sigmoid cross-entropy for CuratorNet.

Notice that in this article we used a pre-trained ResNet \cite{he2016deep} to obtain the image visual features, but the model could use other CNNs such as AlexNet \cite{krizhevsky2012imagenet}, VGG \cite{simonyan2014very},  etc. We chose ResNet since it has performed the best in transfer learning tasks \cite{kornblith2018transfer,delrio2018transfer}. 


\subsection{Data Sampling for Training}

The original BPR article \cite{rendle2009bpr} suggests the creation of training triples $(u, i_{+}, j_{-})$ simply by, given a user $u$, randomly sampling a positive element $i_{+}$ among those consumed, as well as sampling a negative feedback element $j_{-}$ among those not consumed. However, eventual research has shown that there are more effective ways to create these training triples \cite{ding2018improved}. In our case, we define some guidelines to sample triples for the training set based on analyses from previous studies indicating features which provide signals of user preference. For instance, Messina et al. \cite{messina2018content} showed that people are very likely to buy several artworks with similar visual themes, as well as from the same artist, then we used \textit{visual clusters} and \textit{user's favorite artist} to set some of these sampling guidelines.

\begin{figure}[t!h]
    \centering
    \includegraphics[scale=0.4]{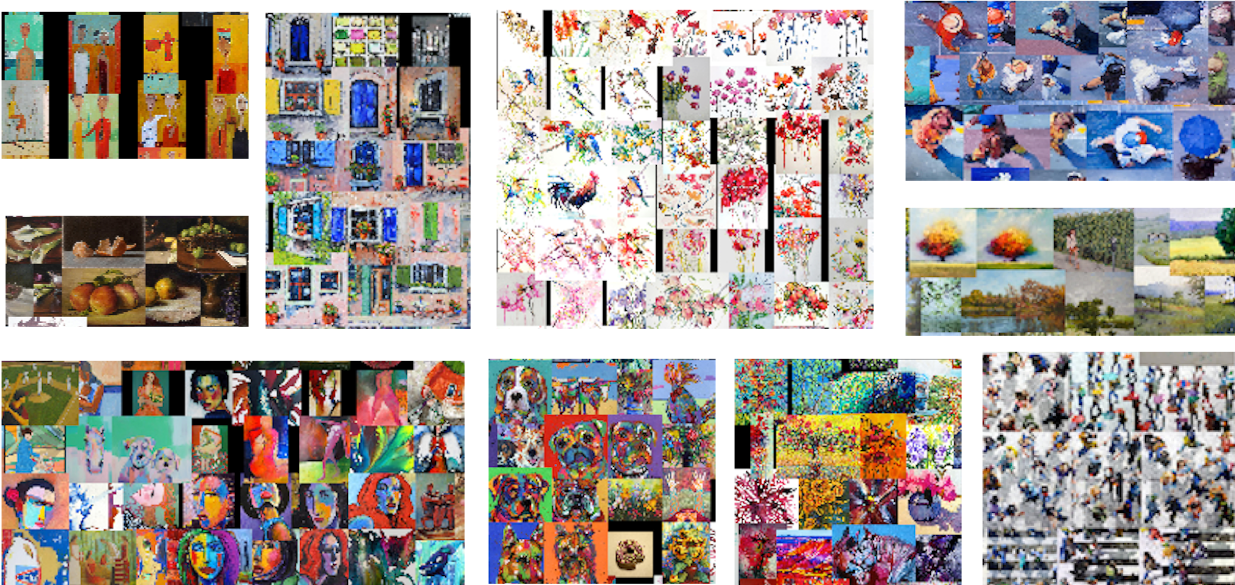}
    \caption{Examples of visual clusters automatically generated to sample triples for the training set.}
    \label{fig:embeddings}
\end{figure}

\textbf{Creating Visual Clusters}. Some of the sampling guidelines are based on visual similarity of the items, and although we have some metadata for the images in the dataset, there is a significant number of missing values: only 45\% of the images have information about subject (e.g., architecture, nature, travel) and 53\% about style (e.g., abstract, surrealism, pop art). For this reason, we conduct a clustering of images based on their visual representation, in such a way that items with visual embeddings that are too similar will not be used to sample positive/negative pairs $(i_{+}, j_{-})$. To obtain these visual clusters, we followed the following procedure: (i) Conduct a Principal Component Analysis to reduce the dimensionality of images embedding vectors from $\mathbb{R}^{2,048}$ to $\mathbb{R}^{200}$, (ii) perform k-means clustering with 100 clusters. We conducted k-means clustering 20 times and for each time we calculated the Silhouette coefficient \cite{rousseeuw1987silhouettes} (an intrinsic metric of clustering quality), so we kept the clustering resulting with the highest Silhouette value. Finally, (iii) we assign each image the label of its respective visual cluster. Samples of our clusters in a 2-dimensional projection map of images, built with the UMAP method \cite{mcinnes2018umap}, can be seen in Figure \ref{fig:embeddings}.

\textbf{Guidelines for sampling triples}.  We generate the training set $\mathcal{D_S}$ as the union of multiple disjoint\footnote{Theoretically, these training sets are not perfectly disjoint, but in practice we hash all training triples and make sure no two training triples have the same hash. This prevents duplicates from being added to the final training set.} training sets, each one generated with a different strategy in mind. These strategies and their corresponding training sets are: 


\begin{enumerate}


    \item 
    Removing item from purchase basket, and predicting this missing item.

    \item 
    Sort items purchased sequentially, and then predict next purchase in basket.
        
    \item 
    Recommending visually similar artworks from the favorite artists of a user.
    
    
    \item 
    Recommending profile items from the same user profile.
    
    \item 
    Create an artificial user profile of a single item purchased, and recommending profile items given this artificially created user profile.
    
    \item 
    Create artificial profile with a single item, then recommend visually similar items from the same artist.

\end{enumerate}

Finally, the training set $\mathcal{D_S}$ is formally defined as:

\begin{equation}
    \mathcal{D_S} = \bigcup\limits_{i=1}^{6} \mathcal{D_S}^i
\end{equation}


In practice, we uniformly sample about 10 million training triples, distributed uniformly among the six training sets $\mathcal{D_S}^i$ . Likewise, we sample about 300,000 validation triples. To avoid sampling identical triples, we hash them and compare the hashes to check for potential collisions. Before sampling the training and validation sets, we hide the last purchase basket of each user, using them later on for testing. 

\section{Experiments}
\label{sec:experiments}

\subsection{Datasets}

For our experiments we used a dataset where the user preference is in the form of purchases over physical art (painting and pictures). This private dataset was collected and shared by an online art store. 
The dataset consists of $2,378$ users, $6,040$ items (paintings and photographs) and $5,336$ purchases. On average, each user bought 2-3 items. One important aspect of this dataset is that paintings are one-of-a-kind, i.e., there is a single instance of each item and once it is purchased, is removed from the inventory. Since most of the items in the dataset are one-of-a-kind paintings (78\%) and most purchase transactions have been made over these items (81.7\%) a method relying on collaborative filtering model might suffer in performance, since user co-purchases are only possible on photographs. Another notable aspect in the dataset is that each item has a single creator (artist). In this dataset there are 573 artists, who have uploaded 10.54 items in average to the online art store. 

The dataset\footnote{\url{https://drive.google.com/drive/folders/1Dk7_BRNtN_IL8r64xAo6GdOYEycivtLy}} with transaction tuples (user, item), as well as the tuples used for testing (the last purchase of each user with at least two purchases) are available for replicating our results as well as for training other models.  Due to copyright restrictions we cannot share the original image files, but we share the embeddings of the images obtained with ResNet50 \cite{he2016deep}.



\subsection{Evaluation Methodology}

In order to build and test the models, we split the data into train, validation and test sets. To make sure that we could make recommendations for all cases in the test set, and thus make a fair comparison among recommendation methods, we check that every user considered in the test set was also present in the training set. All baseline methods were trained on the training set with hyperparameters tuned with the validation set.


\begin{table*}[t!hbp]
    \centering

\caption{Results for all methods, sorted by AUC performance. The top five results are highlighted for each metric. For reference, the bottom row presents a random recommender, while the top row presents results of a perfect Oracle.}

\begin{tabular}{llrrrrrrr}
\toprule
 Method &  $\lambda$ (L2 Reg.)  &    AUC &   R@20 &   P@20 &   nDCG@20 &   R@100 &   P@100 &   nDCG@100 \\
\hline
 Oracle    & --                 & \textbf{1.0000}  & \textbf{1.0000}  & \textbf{.0655}  & \textbf{1.0000}  & \textbf{1.0000}  & \textbf{.0131}  & \textbf{1.0000}  \\
 \midrule 
 CuratorNet & .0001         & \textbf{.7204}  & \textbf{.1683}  & \textbf{.0106}  & \textbf{.0966}  & \textbf{.3200}  & \textbf{.0040}  & \textbf{.1246}  \\
 CuratorNet & .001          & \textbf{.7177}  & \textbf{.1566}  & \textbf{.0094}  & \textbf{.0895}  & \textbf{.2937}  & \textbf{.0037}  & \textbf{.1160}  \\
 VisRank  &    --           & \textbf{.7151}  & \textbf{.1521}  & \textbf{.0093}  & \textbf{.0956}  & \textbf{.2765}  & \textbf{.0034}  & \textbf{.1195}  \\
 CuratorNet & 0             & \textbf{.7131}  & \textbf{.1689}  & \textbf{.0100}  & \textbf{.0977}  & \textbf{.3048}  & \textbf{.0038}  & \textbf{.1239}  \\
 CuratorNet & .01           & .7125 & .1235  & .0075  & .0635  & .2548  & .0032  & .0904  \\
 VBPR & .0001               & .6641 & .1368  & .0081  & .0728  & .2399  & .0030  & .0923  \\
 VBPR & 0                   & .6543 & .1287  & .0078  & .0670  & .2077  & .0026  & .0829  \\
 VBPR & .001                & .6410 & .0830  & .0047  & .0387  & .1948  & .0024  & .0620  \\
 VBPR & .01                 & .5489 & .0101  & .0005  & .0039  & .0506  & .0006  & .0118  \\
 \midrule
 Random   & --              & .4973  & .0103  & .0006  & .0041  & .0322  & .0005  & .0098  \\
\bottomrule
\end{tabular}

    \label{tab:results_paintings_dataset}
\end{table*}


Next, the trained models are used to report performance over different metrics on the test set. For the dataset, the test set consists of the last transaction from every user that purchased at least twice, the rest of previous purchases are used for train and validation.

\textit{Metrics}. To measure the results we used several metrics: AUC (also used in \cite{he2016vbpr,he2016vista,kang2017dvbpr}), normalized discounted cumulative gain (nDCG@k)\cite{jarvelin2002cumulated}, as well as Precision@k and Recall@k \cite{cremonesi2010performance}. Although it might seem counter-intuitive, we calculate these metrics for a low (k=20) as well as high values of \textit{k} ($k=100$). Most research on top-k recommendation systems focuses on the very top of the recommendation list, (\textit{k}=5,10,20). However,  Valcarce et al. \cite{valcarce2018robustness} showed that top-k ranking metrics measured at higher values of \textit{k} (\textit{k}=100, 200) are specially robust to biases such as sparsity and popularity biases. The sparsity bias refers to the lack of known relevance for all the user-items pairs, while the popularity bias is the tendency of popular items to receive more user feedback, then missing user-items are \textit{not missing at random}. We are specially interested in preventing popularity bias since we want to recommend not only from the artists that each user is commonly purchasing from. We aim at promoting novelty as well as discovery of relevant art from newcomer artists.

\subsection{Baselines}

The methods used in the evaluation are the following:

\begin{enumerate}[leftmargin=*]
    \item \textbf{CuratorNet}: The method described in this paper. We also test it with four regularization values for $\lambda = \{0, .01, .001, .0001\}$.

    \item \textbf{VBPR} \cite{he2016vbpr}: The state-of-the-art. We used the same embedding size as in CuratorNet (200), we optimized it until converge in the training set and we also tested the four regularization values for $\lambda = \{0, .01, .001, .0001\}$. 
    
    \item \textbf{VisRank} \cite{kang2017dvbpr,messina2018content}: This is a simple memory-based content filtering method that ranks a candidate painting $i$ for a user $u$ based on the maximum cosine similarity with some existing item in the user profile $j \in P_u$ i.e. 
    \begin{equation}
        score(u,i) = max_{j \in P_u} {cosine(i,j)}
        \label{eq:visrank}
    \end{equation}

    
\end{enumerate}

\begin{figure}[t!]
    \centering
    \includegraphics[scale=0.42,trim={0 0 0 0},clip]{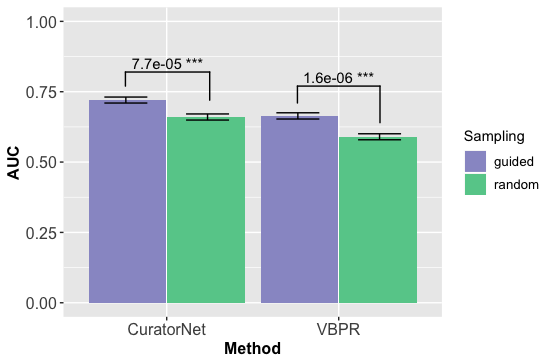}
    \vspace{0mm}
    \caption{The sampling guidelines had a positive effect on AUC compared to random negative sampling for building the BPR training set.
    }
    \label{fig:guidelines}
\vspace{0mm}
\end{figure}   

\section{Results and Discussion}
\label{sec:results-discussion}

In Table \ref{tab:results_paintings_dataset}, we can see the results comparing all methods. As reference, at the top rows we present an oracle (perfect ranking), and in the bottom row a random recommender. Notice that AUC for a random recommender should be theoretically =0.5 (sorting pairs of items given a user), so the AUC$=.4973$ serves as a check. In terms of AUC, Recall@100, and Precision@100 CuratorNet with a small regularization ($\lambda=.0001$) is the top model among other methods. We highlight the following points from these results:
\begin{itemize}[leftmargin=*]
    \item CuratorNet, with a small regularization $\lambda=.0001$, outperforms the other methods in five metrics (AUC, Precision@20, Recall@100, Precision@100 and nDCG@100), while it stands second in Recall@20 and nDCG@20 against the non-regularized version of CuratorNet. This implies that CuratorNet overall ranks very well at top positions, and is specially robust against sparsity and popularity bias \cite{valcarce2018robustness}. In addition, CuratorNet seems robust to changes in the regularization hyperparameter.

    \item Compared to VBPR, CuratorNet is better in all seven metrics (AUC, Precision@20, Recall@100, Precision@100 and nDCG@100). Notably, it is also more robust to the regularization hyperparameter $\lambda$ than VBPR. We think that this is explained in part due to the characteristics of the dataset: VBPR exploits non-visual co-occurrance patterns, but in our dataset this signal provides a rather small preference information, since almost 80\% are one-of-a-kind items and transactions.
    

    
    \item VisRank presents very competitive results, specially in terms of AUC, nDCG@20 and nDCG@100, performing better than VBPR in this high one-of-a-kind dataset. However, CuratorNet performs better than VisRank in all metrics. This provides evidence that the model-based approach of CuratorNet that aggregates user preferences into a single embedding is a better approach than the heuristic-based scoring of VisRank.
\end{itemize}

\subsection{Effect of Sampling Guidelines}

We studied the effect of using our sampling guidelines for building the training set $\mathcal{D_S}$ compared to the traditional BPR setting where negative samples $j$ are sampled uniformly at random from the set of unobserved items by the user, i.e., $I \setminus I_u^+$. In the case of CuratorNet we use all six sampling guidelines ($\mathcal{D_S}^1-\mathcal{D_S}^6$), while in VBPR we only used two sampling guidelines ($\mathcal{D_S}^3$ and $\mathcal{D_S}^4$), since VBPR has no notion of session or purchase baskets in its original formulation, and it has more parameters than CuratorNet to model collaborative non-visual latent preferences. We tested AUC in both CuratorNet and VBPR, under their best performance with regularization parameter $\lambda$, with and without our sampling guidelines. Notice that results in Table \ref{tab:results_paintings_dataset} all consider the use of our sampling guidelines. After conducting pairwise t-tests, we found a significant improvement in CuratorNet and VBPR, as shown in Figure \ref{fig:guidelines}. CuratorNet with sampling guidelines (AUC=$.7204$) had a significant improvement over CuratorNet with random negative sampling (AUC=$.6602$), $p=7.7\cdot10^{-5}$. Likewise, VBPR with guidelines (AUC=$.6641$) had a significant improvement compared with VBPR with random sampling (AUC=$.5899$), $p=1.6\cdot10^{-6}$. With this result, we conclude that the proposed sampling guidelines help in selecting better triplets for more effective learning in our art image recommendation setting.

\section{Conclusion}
\label{sec:conclusion}

In this article we have introduced CuratorNet, an art image recommender system based on neural networks. The learning model of CuratorNet is inspired by VBPR \cite{he2016vbpr}, but it incorporates some additional aspects such as layers with shared weights and it works specially well in situations of one-of-a-kind items, i.e., items which disappear from the inventory once consumed, making difficult to user traditional collaborative filtering. Notice that an important contribution of this article are the data shared, since we could not find on the internet any other dataset of user transactions over physical paintings. We have anonymized the user and item IDs and we have provided ResNet visual embeddings to help other researchers building and validating models with these data.  

Our model outperforms state-of-the-art VBPR as well as other simple but strong baselines such as VisRank \cite{kang2017dvbpr,messina2018content}. We also introduce a series of guidelines for sampling triples for the BPR training set, and we show significant improvements in performance of both CuratorNet and VBPR versus traditional random sampling for negative instances.

\textbf{Future Work}. Among our ideas for future work, we will test our neural architecture using end-to-end-learning, in a similar fashion than \cite{kang2017dvbpr} who used a light model called CNN-F to replace the pre-trained AlexNet visual embeddings. Another idea we will test is to create explanations for our recommendations based on low-level (textures) and high level (objects) visual features which some recent research are able to identify from CNNs, such as the Network Dissection approach by Bau et al. \cite{bau2017network}. Also, we will explore ideas from the research on image style transfer \cite{ghiasi2017exploring,gatys2016image}, which might help us to identify styles and then use this information as context to produce style-aware recommendations. Another interesting idea for future work is integrating multitask learning in our framework, such as the recently published paper on the newest Youtube recommender \cite{Zhao2019YT}. Finally, from a methodological point-of-view, we will test other datasets with \textit{likes} rather than \textit{purchases}, since we aim at understanding how the model will behave under a different type of user relevance feedback.

\begin{acks}
This work has been supported by the Millennium Institute for Foundational Research on Data (IMFD) and by the Chilean research agency ANID, FONDECYT grant 1191791.
\end{acks}

%
\bibliographystyle{ACM-Reference-Format}
\bibliography{sample-base}

\end{document}